\documentclass[titlepage,12pt,twoside]{article}
\usepackage{amssymb,epsfig,pslatex}
\usepackage{cite}  
\usepackage{float}       
\usepackage{wrapfig}  
\usepackage{hhline} 
\usepackage{dcolumn} 
\usepackage{dsfont}
\restylefloat{figure} 

\def\ttbar{{t \bar t}} 

\newcommand{\rmi}{\rm i}
\newcommand{\Or}{{\cal O}}

\newcommand{\be}{\begin{equation}}
\newcommand{\ee}{\end{equation}}
\newcommand{\bea}{\begin{eqnarray}}
\newcommand{\eea}{\end{eqnarray}}

\newcommand{\st}{{\mathbf S}_t}

\newcommand{\bfig}{\begin{figure}}
\newcommand{\efig}{\end{figure}}
\newcommand{\bc}{\begin{center}}
\newcommand{\ec}{\end{center}}

\makeatletter
\def\@sect#1#2#3#4#5#6[#7]#8{\ifnum #2>\c@secnumdepth
  \def\@svsec{}\else
  \refstepcounter{#1}\edef\@svsec{\csname the#1\endcsname.\hskip0.5em}\fi
  \@tempskipa #5\relax
  \ifdim \@tempskipa>\z@
    \begingroup
      #6\relax
      \@hangfrom{\hskip #3\relax\@svsec}{\interlinepenalty \@M #8\par}%
    \endgroup
    \csname #1mark\endcsname{#7}\addcontentsline
      {toc}{#1}{\ifnum #2>\c@secnumdepth \else
        \protect\numberline{\csname the#1\endcsname}\fi #7}%
  \else
    \def\@svsechd{#6\hskip #3\@svsec #8\csname #1mark\endcsname
      {#7}\addcontentsline{toc}{#1}{\ifnum #2>\c@secnumdepth \else
        \protect\numberline{\csname the#1\endcsname}\fi #7}}%
  \fi \@xsect{#5}}
\@addtoreset{equation}{section}
\makeatother
\renewcommand\thesection{\arabic{section}}
\renewcommand\theequation{\ifnum \value{section}>0
 \thesection.\arabic{equation}%
\else
\arabic{equation}%
\fi}

\renewcommand{\thefootnote}{\small\fnsymbol{footnote}}

\begin{document}
\begin{titlepage}
  \begin{flushright}
    PITHA 08/30 
  \end{flushright}        
\vspace{0.01cm}
\begin{center}
{\LARGE {\bf The Top Quark Decay Vertex \\ in Standard Model Extensions}}  \\
\vspace{2cm}
{\large{\bf Werner Bernreuther\footnote{Email:
{\tt breuther@physik.rwth-aachen.de}},
 Patrick Gonz\'alez\footnote{Email:
{\tt gonzalez@physik.rwth-aachen.de}},
 Martin Wiebusch\footnote{Email: {\tt mwiebusch@physik.rwth-aachen.de}} 
}}
\par\vspace{1cm}
Institut f\"ur Theoretische Physik, RWTH Aachen University, 52056 Aachen, Germany\\
\par\vspace{1cm}
{\bf Abstract}\\
\parbox[t]{\textwidth}
{New physics interactions can affect the strength and structure of the
  $tbW$ vertex. We investigate the magnitudes and phases of
  ``anomalous'' contributions to this vertex in a two-Higgs doublet and  
the minimal supersymmetric extension of the standard model, and in  
  a top-color assisted technicolor (TC2) model.  While the magnitudes  
  of the anomalous couplings remain below 1 percent  in the
  first two models, TC2 interactions can reduce the left-chiral
  coupling $f_L$ by several percent. 
}
\end{center}
\vspace*{2cm}

PACS number(s): 11.30.Er, 12.60.-i,  14.65.Ha  
\end{titlepage}
%
%
\setcounter{footnote}{0}
\renewcommand{\thefootnote}{\arabic{footnote}}
\setcounter{page}{1}
\section{Introduction}
\label{Intro} \quad
The decays of top quarks are a direct and sensitive probe of the
fundamental interactions at energy scales of a few hundred GeV.
 So far all data from the Tevatron are compatible with the predictions from
 the Standard Model (SM).
 Only the decay mode $t\to Wb$ has been detected \cite{Varnes:2008tc},
which is predicted
by the SM to completely dominate the top-decay rate. 
SM extensions (BSM) suggest that exotic decay modes 
of the top quark may exist, with branching ratios
 being observable at the Tevatron or eventually
 at the Large Hadron Collider (LHC). Well-known examples
 include the decay into a relatively light
charged Higgs boson, $t\to b H^+$, or into a light top-squark
 and the lightest neutralino, $t\to {\tilde t}_1 {\tilde \chi}_1$,
 which is possible in the (minimal) supersymmetric SM extension (MSSM).
 (See, for instance, the 
reviews \cite{Chakraborty:2003iw,Wagner:2005jh,Quadt:2006jk,Bernreuther:2008ju}).

BSM physics affecting top quarks need not lead to new decay modes,
 because of kinematic obstructions. In any case, it should 
leave its mark on the strength and structure of the $tbW$ vertex.
Parameterizing BSM contributions to the vertex by ``anomalous''
couplings, one can obtain direct information on these couplings
 from the fractions $F_{0,\mp}$ of $t \to b W^+$ decays with
 $W$-boson helicity $\lambda_W = 0,\mp 1$, and 
from single top-quark production. While the sensitivity 
 to BSM effects  on
 $F_{0,\mp}^{exp}$ 
 \cite{Abulencia:2006ei,Abulencia:2006iy,Abazov:2006hb,Abazov:2007ve}
 and on single-top-quark 
production \cite{Abazov:2006gd,Abazov:2008kt,Abazov:2008sz,Aaltonen:2008sy}
from the Tevatron is rather modest,
 one expects that these anomalous form factors can be determined
 quite precisely at the LHC 
 \cite{Hubaut:2005er,Tsuno:2005qb,AguilarSaavedra:2006fy,AguilarSaavedra:2007rs,AguilarSaavedra:2008gt}.

On the theoretical side, there have been a number of investigations on
 new physics contributions to $t\to Wb$ decay, including two-Higgs
 doublet models (2HDM) \cite{Grzadkowski:1991nj,Denner:1992vz}, 
the minimal supersymmetric extension of the standard model (MSSM)  
\cite{Li:1992ga,Yang:1993ra,Garcia:1993rq,Dabelstein:1995jt,Brandenburg:2002xa,Cao:2003yk},
 top-color assisted
 technicolor models (TC2) \cite{Wang:2005ra}, and Little Higgs models
 \cite{Cao:2006wk,Berger:2005ht,Penunuri:2008pb}. 
  In these papers, the corrections to the decay rate and, in some cases, also to the
 helicity fractions were analyzed.

In this paper we investigate  the $tbW$ vertex in a number of SM
extensions,  i.e.,  we compute the induced anomalous 
charged current couplings, revisiting and extending previous results
in the literature.  Our primary concern is to investigate whether BSM
models predict large enough values for these couplings to be detectable at the LHC.  
In Section~\ref{secaom} we specify our convention for these form
factors  and
recapitulate their $CP$ transformation properties. We briefly review
 what is presently known  from experiments about these couplings, and recall
 the sensitivities with which they are expected to be measurable at
 the LHC. In Section~\ref{sesffsmext} we compute these anomalous
 couplings at the 1-loop level in the type-II two-Higgs doublet model, in the
 MSSM, and in a TC2 model. For completeness we recapitulate 
 also some results from Little Higgs
models. Apart from determining the magnitude of these form factors, we
investigate also their phases. In the context of these models this
 provides a check of the assumption, often made in simulation studies
  \cite{Hubaut:2005er,Tsuno:2005qb,AguilarSaavedra:2007rs},
 that these couplings are real to a good approximation.
 We conclude in Section~\ref{sec-conc}.

\section{The $tbW$ vertex: status and expectations}
\label{secaom} \quad
As is well-known, 
a model-independent analysis of the structure of the
 $tbW$ vertex can be made using a form-factor decomposition.
The amplitude ${\cal M}_{tbW}$ of the decay $t(p) \to b(k) \, W^{+}(q)$,
where all particles are on-shell, 
can be decomposed in terms of four form factors:
\be
{\cal M}_{tbW^+} = - \frac{g_W}{\sqrt 2} 
\epsilon^{\mu*} \, {\bar u}_b \left[(V^*_{tb} +f_L) \gamma_\mu P_L +
f_R\gamma_\mu P_R + {\rmi} \sigma_{\mu\nu}  q^\nu 
(\frac{g_L }{m_W} P_L  + \frac{g_R }{m_W} P_R)
\right] u_t \, ,
\label{eq-ffdecomp} 
\ee
with $P_{L,R}=(1\mp\gamma_5)/2$. Here $V_{tb}$ is the Cabibbo-Kobayashi-Maskawa (CKM)
 matrix element in the three-generation  SM, and $p$, $k$, and
$q=p-k$  denote the four-momenta of the 
 $t$  and $b$ quark  and the  $W$ boson, respectively.
 The two chirality conserving 
and  flipping form factors
$f_{L,R}$ and  $g_{L,R}$, respectively,
 are  dimensionless (complex) functions of 
$q^2$. If the $W$ boson is off-shell,
  two additional form factors appear in the matrix element
(\ref{eq-ffdecomp}). However, they
do not contribute to the matrix element of 
$t \to b f_1 {\bar f}_2$ in the limit of vanishing fermion masses $m_{f_{1,2}}$. 

The parameterization in (\ref{eq-ffdecomp})
is chosen in such a way that non-zero values of  $f_{L,R}$ and  $g_{L,R}$
signify deviations from the structure of the tree-level Born vertex.
They are generated by SM loop corrections and, possibly, by new
physics interactions. 
In the SM and in SM extensions which correspond to renormalizable theories,
 $f_{L,R}\neq 0$ can appear at tree-level while $g_{L,R}\neq 0$ must
be loop-induced. Notice that a significant deviation of
$|V_{tb}|_{exp}$ from  $\sim 0.99$ is a possibility which is not yet
 experimentally excluded  \cite{Alwall:2006bx}. In our parameterization
 (\ref{eq-ffdecomp}) this would imply a sizeable coupling $f_L \neq 0$. 

The form factors are gauge-invariant but are not, in general,
 infrared-finite. They should be used to parameterize only new
``infrared safe'' short-distance contributions to the $tbW$ vertex,
caused for instance by the exchange of new heavy virtual particles. 
A search for anomalous couplings  in $t \to b W$ decay-data 
should use the following matrix elements: a) The SM decay distributions
$d\Gamma_{SM}$ including radiative corrections, which are presently known
to NLO in the gauge couplings, that is, for $t \to W b$, $ t\to Wb \gamma$,
and $t\to bW g$. b) One adds to  $d\Gamma_{SM}$ the contributions $d\Gamma_{BSM}$ linear
in the ``anomalous''  form factors  $f_{L,R}$ and $g_{L,R}$, which are
generated by the  interference of (\ref{eq-ffdecomp}) with the SM Born
amplitude, and possibly also the terms bilinear\footnote{Within a
  specific SM extension, these linear and bilinear terms must, of
  course, be incorporated consistently in a perturbative  calculation.}
 in $f_{L,R}$, $g_{L,R}$.
  In the following we use the convention that
 $f_{L,R}$ and $g_{L,R}$ parameterize only the new physics
 contributions to $t\to bW$.  

The amplitude of the charge-conjugate decay 
${\bar t}({\bar p}) \to {\bar b}({\bar k}) \, W^{-}({\bar q})$
 has the general structure:
\be
{\cal M}_{{\bar t}{\bar b}W^-} = - \frac{g_W}{\sqrt 2} 
\epsilon^{\mu*} \, {\bar{\rm v}}_t \left[(V_{tb} +f'_L) \gamma_\mu P_L +
f'_R \gamma_\mu P_R + {\rmi} \sigma_{\mu\nu}  {\bar q}^\nu 
(\frac{g'_R }{m_W} P_L  + \frac{g'_L }{m_W} P_R)
\right] {\rm v}_b \, . 
\label{eq-ffdecomcc} 
\ee
 $CP$ invariance requires, apart from $V_{tb}$ being real, that
 the form factors in (\ref{eq-ffdecomp}),  (\ref{eq-ffdecomcc})
 satisfy\footnote{In  \cite{Bernreuther:1992be} a different
 convention was used for the amplitude (\ref{eq-ffdecomcc}).}
\cite{Bernreuther:1992be}:
\be
f'_i = f_i \, , \qquad g'_i = g_i \qquad (i = R, L) \, .
\label{cprel}
\ee
CPT invariance implies another useful set of relations.
If absorptive parts of the form factors are neglected, then
\be
f'_i = f^*_i \, , \qquad g'_i = g^*_i \qquad (i = R, L) \, .
\label{cptrel}
\ee 
These relations imply the following: If absorptive parts 
(i.e., final-state interactions) can be neglected, then $CPT$ invariance enforces the real
parts of the form factors to be equal, even if $CP$ is violated:
\be
{\rm Re} \, f'_i = {\rm Re} \, f_i \, ,  \qquad {\rm Re} \, g'_i = {\rm Re} \,
g_i  \qquad
 (i = R, L) \, .
\label{cptabs}
\ee
In this case $CP$ violation induces  non-zero imaginary parts  which
are equal in magnitude but differ in sign, namely, 
\be
{\rm Im} \, f'_i = - {\rm Im} \, f_i \, , \qquad 
{\rm Im} \, g'_i = - {\rm Im} \, g_i \qquad  (i = R, L) \, .
\label{cpvioabs}
\ee
Non-zero absorptive parts of the decay
amplitude also lead to imaginary parts of the form factors. If $CP$ is
 conserved they are equal in  magnitude and sign. If $CP$ is broken
 then $CP$-violating absorptive parts of the decay amplitude can
 contribute to the real parts of the form factors and violate
(\ref{cptabs}). Suffice it to say that $CP$ violation at an observable
level in the
 CKM-allowed decay $t\to bW \to b q {\bar q}'$ would be a clear sign of a non-SM
 $CP$-violating interaction. CKM induced $CP$ violation in these decay
 modes is unobservably small, as it is a higher loop effect proportional to 
$J_{CP} = {\rm Im}(V_{tb}V_{cd}V^*_{td}V^*_{cb})
\simeq 3 \times 10^{-5}$. 
%
\begin{table}
\centering
\begin{tabular}{ccccc}
\hline
\hline
  & $f_L$ & $f_R$ & $g_L$ & $g_R$ \\
\hline
${\rm upper \; bound}$ & 0.03& 0.0025& 0.0004 & 0.57 \\
${\rm lower \; bound}$ & $-0.13$ & $-0.0007$ & $-0.0015$ & $-0.15$  \\
\hline
\hline
\end{tabular}
\caption{\label{tab-anobound} Current 95 $\%$ C.L. upper and lower
bounds on the anomalous form factors in the $tbW$ vertex from  
$B({\bar B}\to X_s \gamma)$ \cite{Grzadkowski:2008mf}. Here the form factors
 were assumed to be real.}
\end{table}

For a small $V+A$ admixture to the SM current, energy
and higher-dimensional distributions were computed
in the fashion described above eq. (\ref{eq-ffdecomcc})
   in \cite{Jezabek:1994zv,Bernreuther:2003xj}.
In this case neutrino energy-angular distributions
turn out to be most sensitive to $f_R \neq 0$.
It is important to take the
QCD corrections into account in (future) data analyses, as gluon radiation  can mimic
a small $V+A$ admixture. 

There are indirect constraints on some of the anomalous couplings
from the measured branching ratio  $B({\bar B}\to X_s \gamma)$. 
In particular the constraints on $f_R$ and $g_L$ -- that is,  the
couplings to a right-chiral $b$ quark --
 are tight, as the contributions of these couplings to the respective
 $B$-meson decay  amplitude are enhanced by
a factor $m_t/m_b$  \cite{Fujikawa:1993zu,Cho:1993zb}.
A recent analysis \cite{Grzadkowski:2008mf}
arrives at the bounds given in table \ref{tab-anobound}. (For earlier work,
 see \cite{Larios:1999au,Burdman:1999fw}.) These bounds were obtained
 by allowing only one coupling to be non-zero at a time. 
Obviously, these bounds are no substitute for direct measurements, as the contributions of different
couplings to this decay rate might cancel among each other or might be
off-set by other new physics contributions.

A basic direct test of
the structure of the $tbW$ vertex is the measurement of the decay fractions 
$F_0=B(t\to b W(\lambda_W =0))$, $F_\mp =B(t\to b W(\lambda_W =\mp 1))$ 
into $W^+$ bosons of helicity $\lambda_W =0,\mp 1$. 
(By definition $F_0 + F_- + F_+ = 1$.) Their SM values are known,
 including the ${\Or}(\alpha_s)$ QCD
and $\Or(\alpha)$ electroweak corrections  \cite{Do:2002ky}, which are,
in fact, small corrections.
For $m_t=172$ GeV, $F_0 = 0.689$, $F_- = 0.310$, $F_+ = 0.001$ in the
SM. The dependence on the anomalous couplings (\ref{eq-ffdecomp}) of
 the helicity fractions $F_{0,\pm}$ can be obtained in straightforward
 fashion. Corresponding expressions can be found, for instance, in
 \cite{Chen:2005vr,AguilarSaavedra:2006fy}. The fractions
 $F_{0,\pm}$ are sensitive only to ratios of couplings. The top width
 $\Gamma_t$ is an observable which is sensitive also to the absolute
 strength of the $tbW$ vertex. However, no
  method is presently known to measure $\Gamma_t$ with reasonable precision at
  a hadron collider. Single top-quark production is the means to get
 a handle on the strength of the $tbW$ vertex, and in particular on
 $f_L$. In the SM, top quarks can be produced singly at hadron
 colliders by $t$-channel reactions ($qb \to q' t$), which are the
 dominant production processes both at the Tevatron and at the LHC,
by $s$-channel reactions ($q {\bar q}' \to t {\bar b}$), and by
associated production $gb \to t W^-$.  Assuming that new
physics effects in these processes reside in the $tbW$ vertex only, one
 can compute the corresponding 
 cross sections in terms of the above anomalous couplings 
  \cite{Boos:1999dd,Chen:2005vr,Cao:2007ea,Najafabadi:2008pb,AguilarSaavedra:2008gt}. 
The single top  processes
may eventually be separately measurable at the LHC. 
The four anomalous
  couplings (\ref{eq-ffdecomp}) may, in principle,  be determined simultaneously
 from the  observables
$F_{0,\pm}$ and the separately measured single-top cross sections, 
if these quantities will be measured with 
 sufficient precision. This should eventually be feasible at the LHC.
 In addition to the helicity fractions, energy and energy-angular 
 distributions in polarized semi- and non-leptonic top-quark decays turn out to
 be  good probes for $f_R$  \cite{Jezabek:1994zv,Bernreuther:2003xj}
  and $g_R$ \cite{Tsuno:2005qb}. (Cf. also \cite{Nelson:1997xd,Nelson:1998pu,Nelson:2000dn}.)

The CDF and D0 experiments at the Tevatron have measured the helicity
fractions from top quark decays in $\ttbar$ events
\cite{Abulencia:2006ei,Abulencia:2006iy,Abazov:2006hb,Abazov:2007ve}.
Furthermore, both the D0 and the CDF experiment have found evidence for single
 top quark production  \cite{Abazov:2006gd,Abazov:2008kt,Aaltonen:2008sy},
 and the  D0 collaboration has recently made a
 search for anomalous couplings \cite{Abazov:2008sz} based on their
 measurement of $\sigma_t$. Both the measured helicity fractions and the D0 and CDF results for
$\sigma_t$ are compatible with SM expectations. The level of precision
with which these observables are presently known does not imply constraints
 on $f_{L,R}$ and $g_{L,R}$ that can compete with the indirect bounds
 of table~\ref{tab-anobound}.

However, future high statistics data on top quark
decays at the LHC will allow direct determination of 
the couplings $f_R, g_L$, and $g_R$ with an accuracy of 
a few percent\footnote{Effects of anomalous couplings in the CKM-suppressed
$tsW$ vertex were studied in  \cite{Lee:2008xr}.}. 
Simulation studies 
analyzed $\ttbar$ production and decay into 
lepton plus jets channels  \cite{Hubaut:2005er,Tsuno:2005qb,AguilarSaavedra:2007rs}
and also  dileptonic channels
\cite{Hubaut:2005er}. Basic observables for determining
the anomalous couplings are the $W$-boson helicity fractions
and associated  forward-backward asymmetries
\cite{AguilarSaavedra:2007rs}.
A double angular distribution in $t$-quark decay was used in \cite{Tsuno:2005qb}.
 These simulation studies assumed  $f_L$ to be zero and
 the other form factors to be real.
The parametric dependence of the observables
on the anomalous couplings yields estimates for the expected
confidence intervals. Assuming that only one non-standard coupling is
nonzero at a time, 
\cite{AguilarSaavedra:2007rs} concludes that values
 of $f_R$, $g_L$, or $g_R$ outside the  following intervals,
\be
f_R\,(2\sigma): \:\: [-0.055, 0.13] \, , \quad 
g_L \,(2\sigma): \: \:  [-0.058,0.026]\, , \quad
g_R \,(2\sigma): \: \: [-0.026, 0.031] \, ,
\label{eq-expanomc}
\ee
should be either detected or excluded at the 2 s.d. level (statistical
and systematic uncertainties).
The analyses of  \cite{Hubaut:2005er} and of  \cite{Tsuno:2005qb} 
 arrived, as far as $g_R$ is
concerned, at a sensitivity level of the same order. Thus the
sensitivity to $g_R$ expected at the LHC is an order of magnitude
better than the current indirect bound given in table
\ref{tab-anobound}.

An analysis  was recently made  \cite{AguilarSaavedra:2008gt} 
 combining single top-quark production and top-quark decay
(from $\ttbar$ events) and expected experimental uncertainties at the
LHC.  (For related work, see
 \cite{Boos:1999dd,Chen:2005vr,Najafabadi:2008pb}.)
 Assuming all form factors to be real, this analysis concludes 
 that a simultaneous four-parameter fit to
 respective data  yields the $1\sigma$ sensitivities 
\bea
f_L\,(1\sigma): \:\: [-0.15, 0.11] \, , \quad 
f_R\,(1\sigma): \:\: [-0.25, 0.25] \, , \nonumber \\
g_L \,(1\sigma): \: \:  [-0.16, 0.16]\, , \quad
g_R \,(1\sigma): \: \: [-0.012, 0.024] \,.
\label{eq-explhc4dim}
\eea
The sensitivity to $g_R$ is essentially as good as the one obtained in
the one-parameter analysis (\ref{eq-expanomc}).

As discussed above, the form factors $f_{L,R}$ and  $g_{L,R}$ can 
 be complex, due to final-state interactions or $CP$
 violation. $CP$-violating contributions to the absorptive parts of
 the form factors are real and invalidate the relations
 (\ref{cptabs}). Well-known observables which require non-zero $CP$-violating
 absorptive parts are asymmetries of partial decay
 rates. For instance\footnote{The asymmetry $A_{CP}$ is $CP$-odd and
   $CPT$-odd, where $T$ refers to a naive $T$ transformation.}:
\begin{equation}
A_{CP}= \frac{\Gamma(t \to b
W^+) - \Gamma({\bar t} \to {\bar b} W^-)}{\Gamma(t \to b
W^+) + \Gamma({\bar t} \to {\bar b} W^-)}   \, .
\label{eq-aspardr}
\end{equation}
An observable which does not require  $CP$-violating absorptive parts
 to be  non-zero is the expectation value of 
 the $T$-odd triple correlation 
${\cal O}= \st \cdot ({\bf \hat p}_{\ell^+}\times {\bf \hat p}_b)$ 
in polarized semileptonic $t$ decay, $t \to b \ell^+ {\nu_\ell}$.
 Here $\st$
denotes the top spin and the hat signifies a unit vector. This
correlation  is sensitive
both to $CP$-invariant absorptive parts
and to $CP$ violation 
 in the $tbW$ vertex. By measuring ${\cal
  O}$ and the  corresponding correlation $\bar{\cal O}$ in $t$ and
$\bar t$ decay, respectively, one can disentangle
 $CP$-violating effects from $CP$-invariant absorptive parts.
 Taking the difference constitutes a 
$CP$-symmetry test in top-quark decay,
 while the sum picks up the $CP$-invariant imaginary parts\footnote{Other $CP$ asymmetries in
top-quark decay were discussed in \cite{Ma:1991ry}. QCD-induced
T-odd asymmetries in $t\to b \ell \nu_\ell g$ were investigated in
\cite{Hagiwara:2007sz}.}.
 One finds that
$\langle{\cal O}\rangle \mp  \langle{\bar{\cal O}}\rangle \propto {\rm
  Im}(g_R \mp g_R')$
\cite{Bernreuther:1992be,Bernreuther:1993xp}. 

At the LHC only single-top-production can provide highly polarized
samples of $t$ and $\bar t$ quarks. They are mostly produced
 by the $t$-channel reactions $q b \to q' t$ (and likewise, $\bar
 t$). It is known \cite{Mahlon:1999gz} that the sample of $t$ quarks
 produced by these processes is almost $100\%$ polarized 
in the direction of the spectator jet $q'$, with the
jet direction determined in the top-quark rest frame. The sample of
$\bar t$ quarks,
produced by the corresponding  $t$-channel  processes,
 has a polarization degree of $-100\%$ with respect to
 the direction of the spectator jet. 
Alternatively,
 one can also use  the direction of one of the
 proton beams as the $t$ and ${\bar t}$ spin axis  \cite{Mahlon:1999gz}.
 These results were obtained
assuming SM interactions in the production process. If
new physics effects result only in modifications of the $tbW$ vertex
according to (\ref{eq-ffdecomp}), it can be shown that 
 the spectator jet directions remain optimal $t$ and $\bar t$ spin
 axes, as long as $|f_{R}|, |g_{L,R}| \ll 1$. 

If LHC experiments  will eventually be  able to collect clean samples of single $t$ and $\bar t$
 events (with charge tagging through their semileptonic decays), then
 one may measure
\be
   A_{\mp} = \langle  {\bf \hat p} \cdot ({\bf \hat p}_{\ell^+}
    \times {\bf \hat p}_b)  \rangle_t
   \, \mp \, 
\langle   {\bf \hat p} \cdot ({\bf \hat p}_{\ell^-}\times
   {\bf \hat p}_{\bar b})  \rangle_{\bar t}  \, ,
\label{cp-singt}
\ee
where ${\bf \hat p}$  is the direction of 
one of the proton beams in the $t$ $({\bar t})$ rest frame. Again, $A_-$ is
 sensitive to possible $CP$-violating effects, in particular to
 $CP$-violating effects in top-quark decay, while $A_+$ is fed
 by $CP$-invariant absorptive parts. (One should keep in mind, however,
 that $A_-$ is not $CP$-odd in the strict sense, as the initial
 $p p$ state is not a $CP$ eigenstate.) The number of single
 $t$ and $\bar t$ quarks that will be produced at the LHC is huge.
 Thus a statistical error   $\delta A_{\mp}^{stat} \sim 1 \%$ in the
  measurement of $A_{\mp}$ seems
 realistic. However, it remains to be investigated whether the 
 systematic uncertainties can reach the same level of precision.
\section{Form factors in SM extensions} 
\label{sesffsmext}
In this section we compute, within the framework of  several
SM extensions,  the new physics contributions to the form
  factors $f_{L,R}$ and  $g_{L,R}$. To begin with, we briefly
 recapitulate the SM results for the top-quark decay width $\Gamma_t$ and 
 differential distributions. The first-order QCD and electroweak
 corrections to $\Gamma_t$ were first computed in \cite{JezKu89}
 and \cite{Denner:1990ns,Eilam:1991iz}, respectively. The
 ${\Or}(\alpha_s^2)$ corrections to $\Gamma_t$ are also known
  \cite{Czarnecki:1998qc,Chet99}. 
  Essentially, only the QCD corrections
  matter and these amount to corrections of about minus $10\%$ with respect
  to the Born width. The first-order SM corrections
  to the helicity fractions $F_{0,\pm}$, which are small,  were determined in
 \cite{Do:2002ky}. Various distributions for (polarized)
 semileptonic and non-leptonic top-quark decay were calculated
 to  ${\Or}(\alpha_s)$ by \cite{Jezabek:1988ja,Czarnecki:1990pe,Fischer:2001gp}
 and \cite{Fischer:2001gp,Brandenburg:2002xr}, respectively.
As to the structure of the 1-loop decay amplitude ${\cal M}_{tbW}$
in the SM: for $m_b \to 0$ the form factors $f_R^{SM}$ and $g_L^{SM}$
vanish, as these couplings accompany Lorentz structures that involve a
right-chiral $b$ quark.
Virtual photon and gluon exchange lead to infrared-divergent 1-loop SM
form factors. These infrared divergences in $d\Gamma$ are cancelled, as usual, by
the contributions from real soft photon and gluon radiation.

Throughout this paper we shall use the following SM parameters:
\bea
  &1/\alpha_{\rm em} = 137.035999679,\quad
   \alpha_s = 0.1176,\quad
   m_Z = 91.1876\,{\rm GeV}, \quad
   m_W = 80.398\,{\rm GeV},&\nonumber\\
  &m_t = 172.6\,{\rm GeV}, \quad
   m_b = 4.79\,{\rm GeV}, \quad
    V_{tb} = 1 .&
\eea
Furthermore, all new physics contributions are calculated relative  to
the 
SM  corrections with a  Higgs-boson  mass of $120\,{\rm GeV}$.

\subsection{Two-Higgs doublet extensions}
\label{sus2hdm}
Two-Higgs doublet models (2HDM) are among the simplest, phenomenologically
viable SM extensions. They are often used as a paradigm for an extended
Higgs-boson sector, which entails new physics effects, for instance Higgs sector $CP$
violation. Within the supersymmetric framework, two Higgs doublet fields are the
minimum requirement for the Higgs sector of a supersymmetric extension of the
SM.

Here we consider a general type-II 2HDM, where the Higgs doublets $\phi_1$ and
$\phi_2$ couple only to right-handed down-type fermions $(d_{iR},\, \ell_{iR})$
and up-type fermions $(u_{iR},\, \nu_{iR})$, respectively.  First we assume the
tree-level Higgs potential to be $CP$-invariant.  The spin-0 physical particle
spectrum of this model consists of two neutral scalar and one pseudoscalar Higgs
boson, $h^0, H^0$ and $A^0$, respectively, and a charged Higgs boson and its
antiparticle, $H^\pm$. Besides the Higgs boson masses $m_{h^0}, m_{H^0},
m_{A^0},$ and $m_{H^+}$ the model involves two more free parameters, which are
commonly defined as angles, namely, $\tan\beta = {\rm v}_2/{\rm v}_1$, where
${\rm v}_{1,2}$ are the vacuum expectation values of the Higgs fields
$\phi_{1,2}$. The angle $\alpha$ describes the mixing of the two $CP$-even
neutral Higgs states which leads to the mass eigenstates $h^0$ and $H^0$.

Experiments that searched for neutral and charged Higgs bosons exclude, when
analyzed in the framework of type-II 2HDM, various regions in the parameter
space of the model; see, for instance, \cite{Abbiendi:2004gn}. The masses of the
neutral Higgs states are, in general, constrained to be not smaller than about
100 GeV \cite{Amsler:2008zz}. The non-observation of $e^+ e^- \to H^+ H^-$ at
LEP2 provides the model-independent lower bound $m_{H^+} > 79.3$ GeV
\cite{Amsler:2008zz}.  However, the rare decay mode $B({\bar B}\to X_s \gamma)$
implies, when analyzed within the type-II 2HDM, the much more stringent bound
$m_{H^+} > 315$ GeV \cite{Gambino:2001ew}, which holds for all values of
$\tan\beta$.
%
\begin{figure}[h]
\begin{center}
\includegraphics[width=14cm]{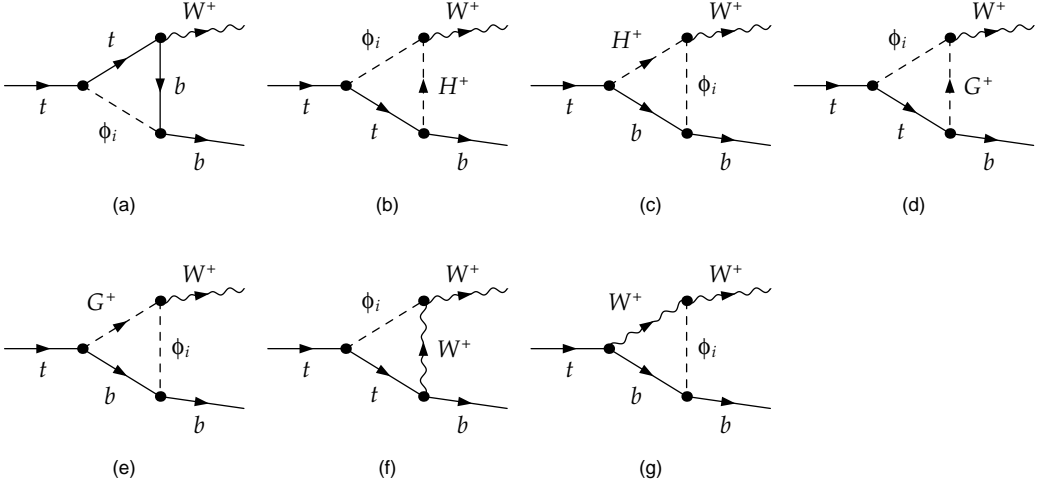}
\end{center}
\caption{Feynman diagrams for the Higgs boson contributions to the $tbW$ vertex
  in the type-II 2HDM. In the case of a 2HDM with $CP$-conserving Higgs
  potential, the symbol $\phi_i$ denotes the neutral bosons $h^0,H^0$ and
  $A^0$. In this case, only the $CP$-even states $\phi_i=h^0, H^0$ contribute in
  Figs. (d), (e), (f), and (g). The self-energy corrections that are involved in the
  renormalization are not shown.}
 \label{fig:higgs}
\end{figure}
The 1-loop Higgs boson contributions to the $t\to bW$ amplitude in the type-II
2HDM are shown in Fig.~\ref{fig:higgs}. The resulting contributions to the decay
width $\Gamma(t\to bW)$ were computed in
\cite{Grzadkowski:1991nj,Denner:1992vz}. Besides using the on-shell
renormalization scheme a suitable choice is to parameterize the lowest order
width in terms of the Fermi constant $G_F$ rather than in terms of the fine
structure constant $\alpha$, in order to avoid large SM corrections
\cite{Denner:1990ns}. The relation between the Born widths in both schemes is
$\Gamma_B(G_F)=\Gamma_B(\alpha)/(1-\Delta r)$, where the well-known quantity
$\Delta r$ summarizes the radiative corrections to muon decay
\cite{Grzadkowski:1991nj,Denner:1992vz}. We also adopt this scheme here and in
the computations within the other models below. We have compared our results of
the non-standard corrections $\delta_{NS}(G_F) =(\Gamma_{NS} -
\Gamma_B(G_F))/\Gamma_B(G_F)$ to the Born width with those of
\cite{Grzadkowski:1991nj,Denner:1992vz} and find agreement. Here and
below, we have used the {\it FeynArts} \cite{Hahn:2000kx,Hahn:2001rv},
 {\it FormCalc}
 \cite{Hahn:1998yk,Hahn:2002vc,Hahn:2004rf,Hahn:2006qw}, 
and {\it LoopTools}\cite{Hahn:1998yk} packages  to perform our calculations.

For this comparison and for the following calculations we have subtracted the
contribution of a Higgs boson with SM couplings from the contributions of the
diagrams of Fig. \ref{fig:higgs}, in order to define the non-standard
corrections. The mass of the SM Higgs boson is chosen to be $120$ GeV.
The Higgs boson masses and $\tan\beta$ of the 2HDM are varied in the following
range which is in accord with experimental constraints: 
\be m_{h^0},\, m_{H^0}, \, m_{A^0}
\geq 120 \, {\rm GeV}, \quad m_{H^+} \geq 320 \, {\rm GeV}, \quad 0.5 \leq \tan
\beta \leq 50 \, .
\label{par2hdm}
\ee

\begin{figure}
\begin{center}
\includegraphics[width=12cm]{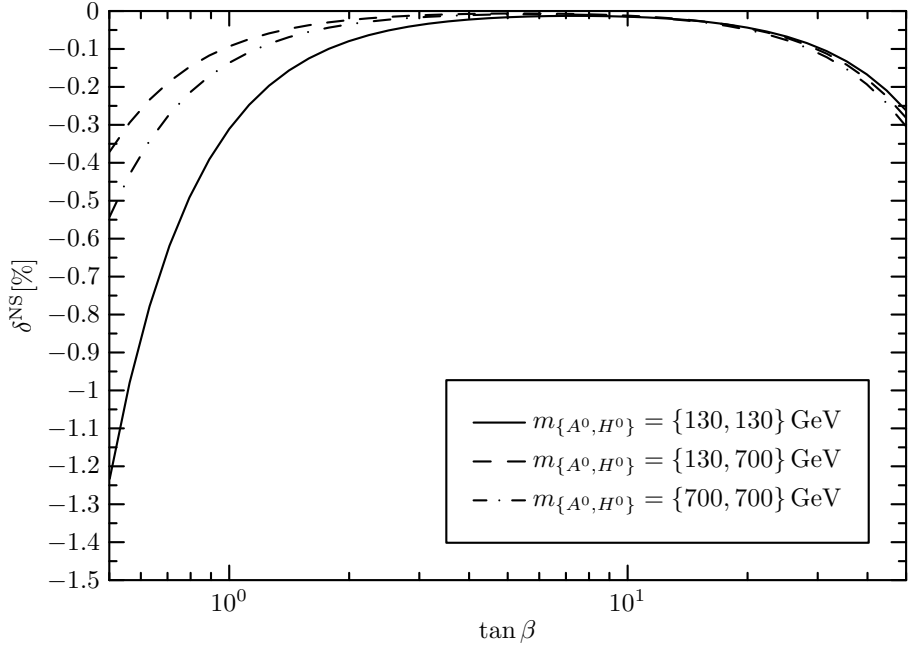}
\end{center}
\caption{The non-standard correction $\delta_{NS}(G_F)$ to the Born width in the
  2HDM as a function of $\tan\beta$ for various Higgs-boson
  masses. Here $m_{h^0}=120$
  GeV, $m_{H^+}=320$ GeV, and the angle $\alpha=\beta-\frac\pi2$.
 \label{fig:delnsthdm}}
 \end{figure}

\begin{figure}
\begin{center}
\includegraphics[width=12cm]{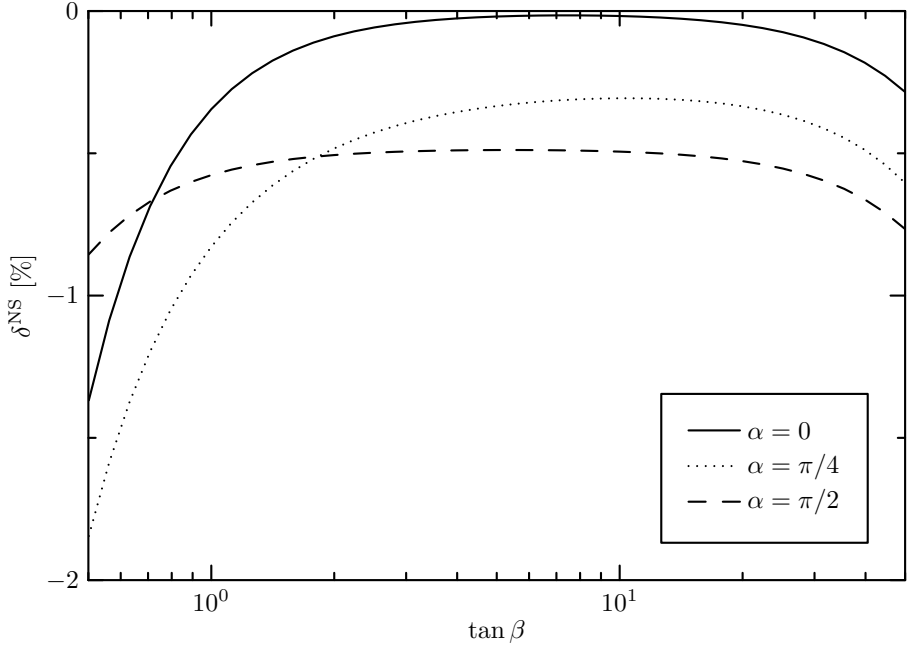}
\end{center}
\caption{The non-standard correction $\delta_{NS}(G_F)$ to the Born width in the
  2HDM as a function of $\tan\beta$ for various (fixed) values of $\alpha$. Here
  $m_{h^0}=120$ GeV, $m_{H^0}=700$ GeV, $m_{A^0}=130$ GeV and $m_{H^+}=320$ GeV.
 \label{fig:delns2dmalpha}}
 \end{figure}

In Fig.~\ref{fig:delnsthdm} the correction $\delta_{NS}(G_F)$ is shown as a
function of $\tan\beta$ for various sets of neutral Higgs masses.  Here the
mixing angle $\alpha$ was put equal to $\beta-\frac\pi2$, which means that the
couplings of the $h^0$ become SM-like. For these parameter sets the corrections
are negative and become largest in magnitude, of the order of $-1\%$, for
$\tan\beta \lesssim 1$, where the Yukawa couplings to the top quark
 are largest. In
Fig.~\ref{fig:delns2dmalpha} we vary $\tan\beta$ while keeping $\alpha$ fixed
and setting $m_{H^0}=700\,{\rm GeV}$ and $m_{A^0}=130\,{\rm GeV}$. We see that
the shape of the curve changes significantly for different values of
$\alpha$. In particular, for $\alpha=0$, corrections of the order of $0.5\%$ are
also possible for intermediate values of $\tan\beta$. However, this strong
$\alpha$ dependence disappears if the two scalar Higgs bosons $h^0$
and $H^0$ have (approximately) the same mass.
%

Let us now determine the anomalous form factors $f_{L,R}$ and $g_{L,R}$. They
are obtained from the decay amplitude, i.e., the contributions of 
Fig.~\ref{fig:higgs}, by appropriate projections. The form factor $f_L$ is
affected by renormalization, while the others are ultraviolet finite.

The bulk of the correction $\delta_{NS}(G_F)$ is due to the renormalized form
factor $f_L$.  By scanning the above parameter range we found the following
generic features: i) $|f_L| \gg |g_R| \gg |f_R|, |g_L|$ and ii) $|{\rm Re} \,
f_L| \gg |{\rm Re} \,g_R| \gg |{\rm Im} \, f_L|, |{\rm Im} \,g_R|$. Hence we
display only the real parts of $f_L$ and $g_R$ in the following. Feature i) can
be qualitatively understood by inspecting the flow of chirality in the diagrams
 of Fig.~\ref{fig:higgs}. As to ii), this follows from the fact that in the
parameter range (\ref{par2hdm}) only diagrams Fig.~\ref{fig:higgs} (e) and (g)
have imaginary parts.

\begin{figure}
\begin{center}
\includegraphics[width=12cm]{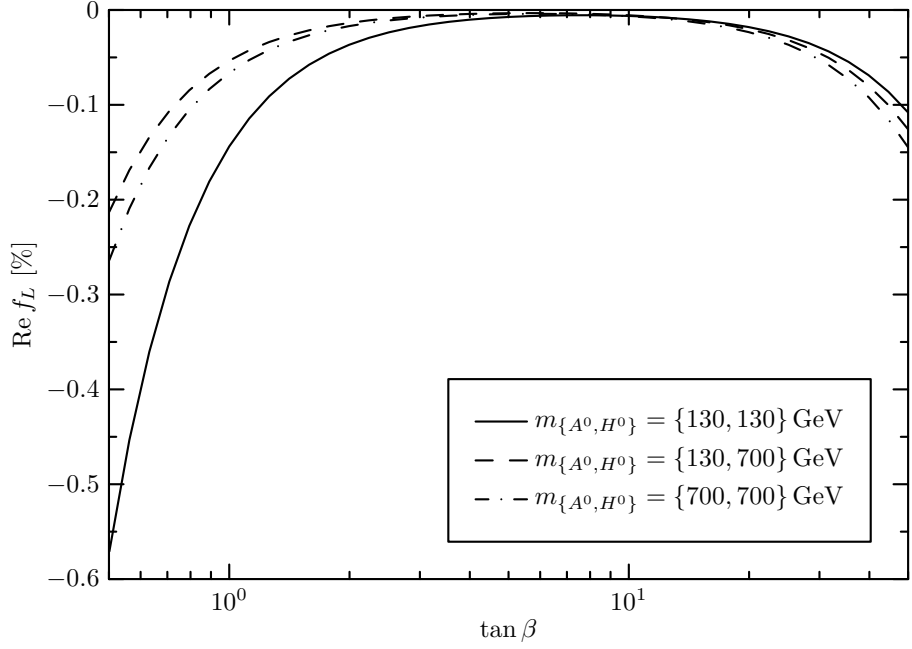}
\end{center}
\caption{The anomalous form factor ${\rm Re} \, f_L$ in the 2HDM as a function
 of $\tan\beta$ for various Higgs-boson masses. The remaining parameters are
 as in  Fig.~\ref{fig:delnsthdm}.
 \label{fig:reflthdm}}
 \end{figure}

In Fig.~\ref{fig:reflthdm} the real part of $f_L$ is shown as a function of
$\tan\beta$ for the same parameter sets as used in Fig.~\ref{fig:delnsthdm}. One
sees that ${\rm Re} \, f_L \approx \delta_{NS}/2$, as expected.

\begin{figure}
\begin{center}
\includegraphics[width=12cm]{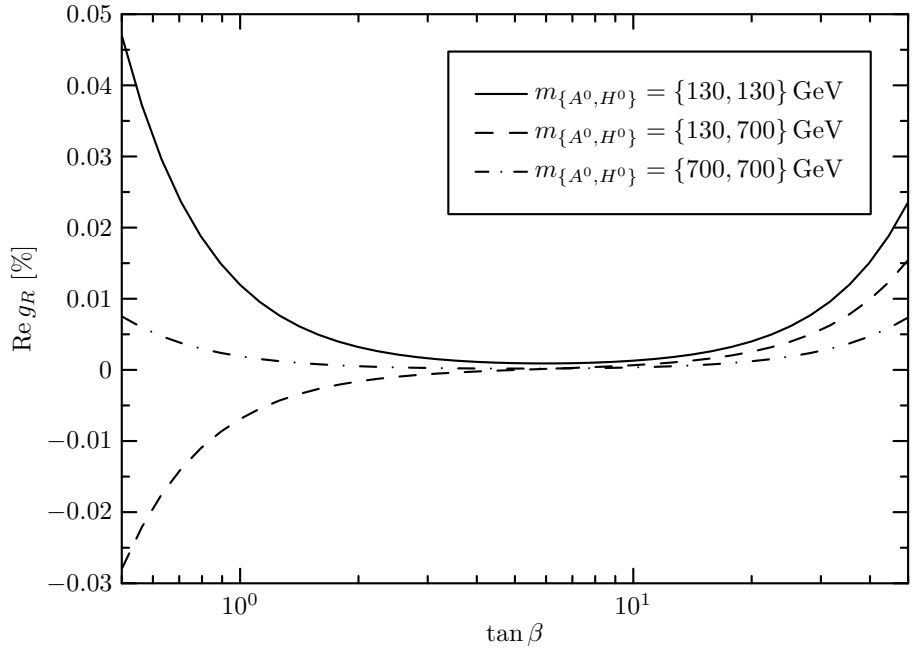}
\end{center}
\caption{The anomalous form factor ${\rm Re} \, g_R$ in the 2HDM as a function
 of $\tan\beta$ for various Higgs-boson masses. The remaining parameters are
 as in Fig.~\ref{fig:delnsthdm}.
 \label{fig:regrthdm}}
 \end{figure}

\begin{figure}
\begin{center}
\includegraphics[width=12cm]{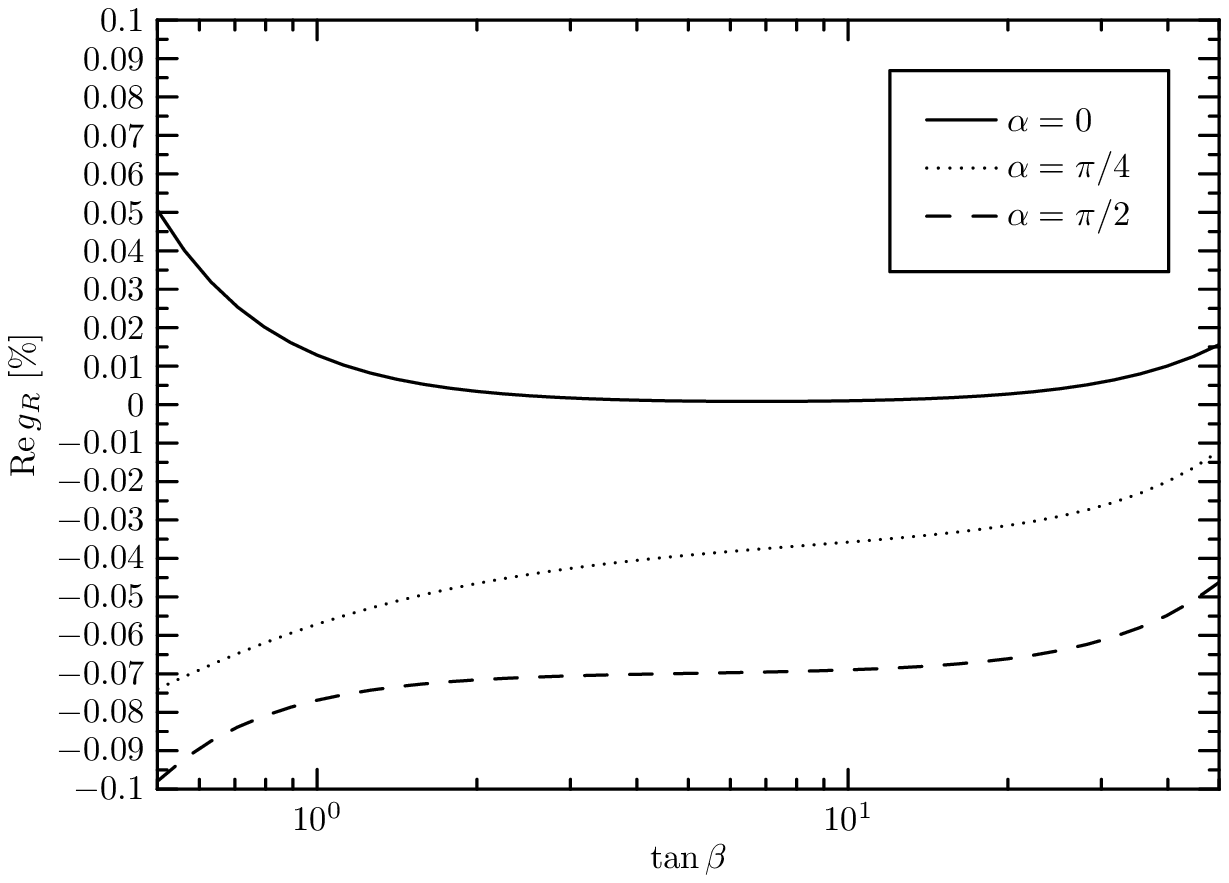}
\end{center}
\caption{The anomalous form factor ${\rm Re} \, g_R$ in the 2HDM as a function
  of $\tan\beta$ for various (fixed) values of $\alpha$. The remaining
  parameters are as in Fig.~\ref{fig:delns2dmalpha}.
 \label{fig:regr2hdmal}}
 \end{figure}

Fig.~\ref{fig:regrthdm} shows the chirality flipping coupling ${\rm Re} \, g_R$
as a function of $\tan\beta$ for the same parameter sets as used in
Fig.~\ref{fig:reflthdm}. This anomalous form factor is one order of 
 magnitude smaller than  ${\rm Re} \, f_L$ -- even for small
(large) values of $\tan\beta$ where the Yukawa couplings of the top (bottom)
quark  become largest.
If the mass difference between the two scalar Higgs bosons
  $h^0$ and $H^0$ is large,
the form factors depend significantly on the mixing angle
$\alpha$. Fig.~\ref{fig:regr2hdmal} shows the $\tan\beta$ dependence of ${\rm
  Re}g_R$ for several fixed values of $\alpha$ and the same parameter sets as
the ones used in Fig.~\ref{fig:delns2dmalpha}. Note that varying $\alpha$ can
change the sign of ${\rm Re}g_R$. The corresponding plot for ${\rm Re}f_L$ is
not shown, since the relation $|{\rm Re}f_L|\gg|{\rm Re}g_R|$ holds for all
values of $\alpha$ and thus ${\rm Re}f_L\approx\delta^{\rm NS}/2$ for all
$\alpha$.
%
\begin{figure}[h]
\begin{center}
\includegraphics[width=7cm]{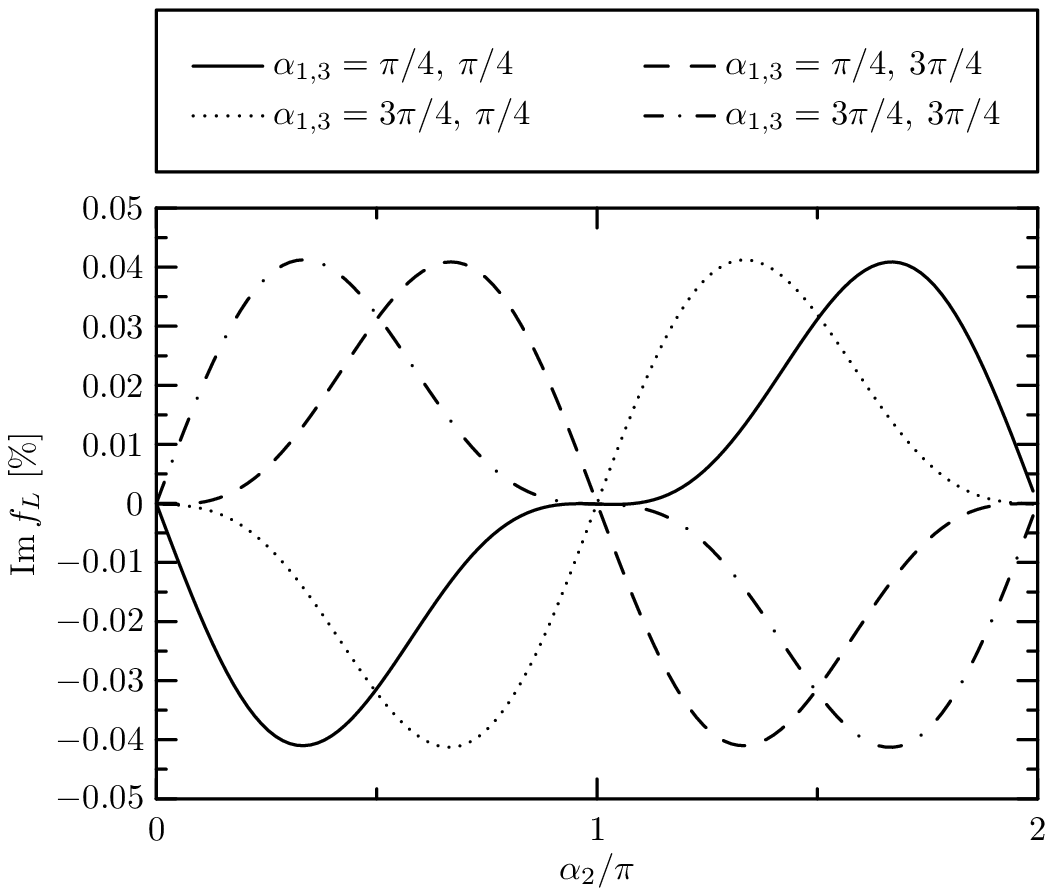}\hspace*{.5cm}
\includegraphics[width=7cm]{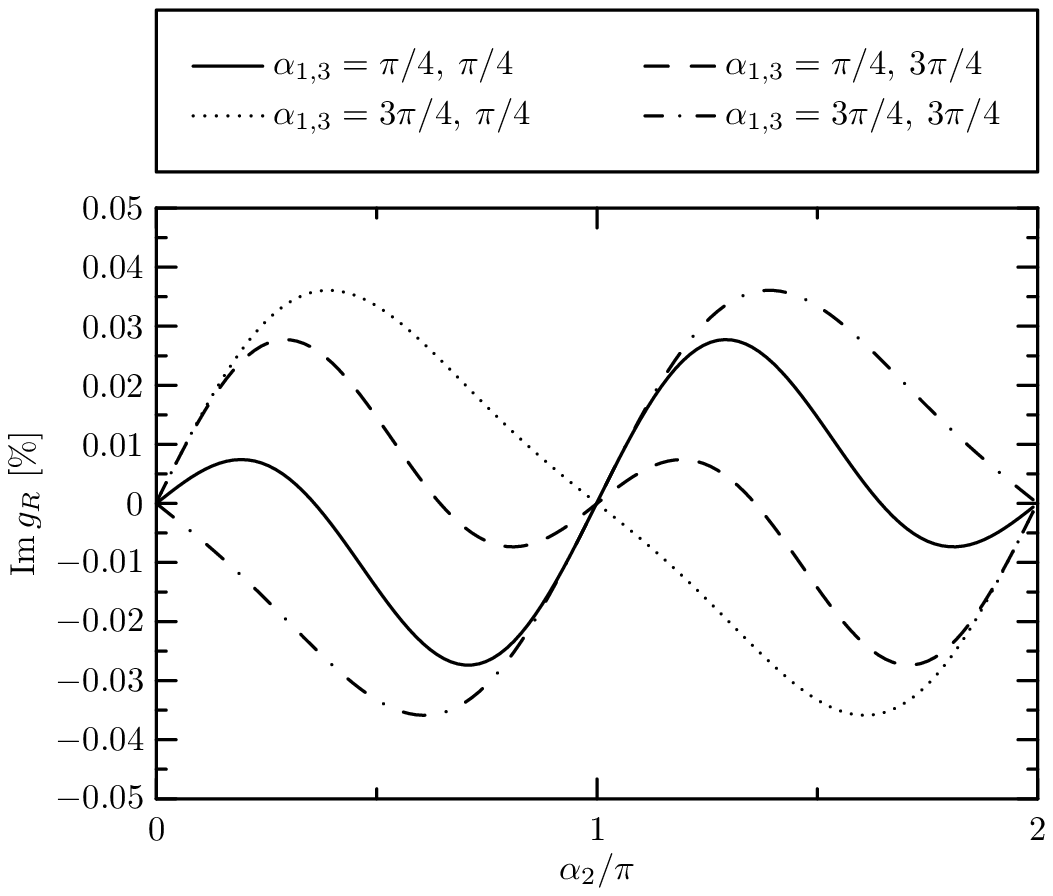}
\end{center}
\caption{Contributions to  ${\rm Im} \, f_L$ (left frame) and
to  ${\rm Im} \, g_R$ (right frame) from $CP$-violating
 neutral Higgs-boson exchange in the 2HDM as a function
 of the mixing angle $\alpha_2$. The remaining parameters are
 given in the text.
 \label{fig:cpthdm}}
 \end{figure}

Neutral Higgs sector $CP$ violation is also possible in 2HDM -- already at Born
level. (Cf., for instance, the recent discussion in
\cite{Maniatis:2007vn}.)
  If the tree-level Higgs potential is not $CP$-invariant,  the states
$h^0$, $H^0$ can mix with $A^0$. Then the 3 neutral physical Higgs-boson mass
eigenstates $\phi_i$ are no longer $CP$ eigenstates; i.e., they couple both to
scalar and pseudoscalar quark and lepton currents.  The $CP$ eigenstates
$\phi'=(h^0, H^0, A^0)$ are related to the mass eigenstates $\phi_i$ by a
real orthogonal matrix $R$, $\phi'_i =R_{ij} \phi_j$. (See, e.g.,
\cite{Bernreuther:1992dz} for the resulting couplings to fermions and weak gauge
bosons). If the neutral Higgs bosons have both scalar and pseudoscalar
couplings to the third-generation quarks then, as discussed in
Section~\ref{secaom}, the form factors acquire imaginary parts -- even if the
decay amplitude has no absorptive part.  With the empirical constraint that
$m_{H^+}>m_t$, only the diagrams Fig.~\ref{fig:higgs} (e) and (g) have
absorptive parts, of which the piece containing the $CP$-even ($CP$-odd)
coupling of the $\phi_i$ to the $b$ quark contributes to the imaginary (real)
part of the form factors. We have discarded these imaginary parts in order to
determine the size of ${\rm Im} f_L$ and ${\rm Im} g_R$ due to $CP$ violation.
Computing ${\rm Im} f_L$ and ${\rm Im} g_R$ by scanning the parameter range
(\ref{par2hdm}) and varying the angles of the mixing matrix $R$, we find that
these couplings do not exceed ${\rm a \; few} \times 10^{-4}$. Effects are
largest if one of the neutral states $\phi_i$ is rather light and the other two
are heavy  and the top-Yukawa couplings of the
light Higgs boson are large. Choosing $m_{\phi_1}=120$ GeV,
$m_{\phi_2}=m_{\phi_3}=700$ GeV, $m_{H^+} = 320$ GeV, $\tan\beta =1$, and
parameterizing the mixing matrix $R=R(\alpha_i)$ in terms of three Euler angles
$\alpha_i$, we have plotted in Fig.~\ref{fig:cpthdm} the $CP$-violating
contributions to ${\rm Im} \, f_L$ and to ${\rm Im} \, g_R$ as a function of
$\alpha_2$, for fixed values of $\alpha_1$ and $\alpha_3$. One sees that $|{\rm
  Im} f_L|\leq 5\times 10^{-4}$ and $|{\rm Im} g_R|\leq 3.5\times 10^{-4}$.

In \cite{Grzadkowski:1992yz,Bernreuther:1993xp,Hasuike:1996tr,Hasuike:1998pn}
the form factor  ${\rm Im} g_R$ was analyzed, with which our results
are in accord. The
 couplings  ${\rm Im} \, f_L$, ${\rm Im} \, g_R$ 
 are too small to cause  observable $CP$-violating effects,
 i.e. a  decay rate asymmetry (\ref{eq-aspardr}) or 
a non-zero triple correlation $A_-$ (cf. eq. (\ref{cp-singt}))
 of the order of $1\%$ or larger.

\subsubsection{The MSSM}
\label{sussusy}

Next we analyze the anomalous form factors within  
 the minimal supersymmetric
 extension  of the SM (MSSM). The Higgs sector of the MSSM corresponds
 to a type-II 2HDM with a $CP$-invariant tree-level Higgs potential. 
Thus the Higgs-boson contributions to the $t\to b W$ decay amplitude are
those depicted in  Fig.~\ref{fig:higgs}. As to the  parameters of the
MSSM Higgs sector: It is well-known that one of the neutral scalar
Higgs bosons, $h^0$, is predicted to be light, $m_{h^0} \lesssim 130$
GeV. Moreover, the lower bound on the mass of the charged Higgs boson
 $H^+$ from 
 $B({\bar B}\to X_s \gamma)$ mentioned above does not apply to the
 MSSM, as the contribution  of $H^+$ to the amplitude of 
 this  decay mode  can be compensated to a large extent by the
 contributions of supersymmetric particles. Thus the mass of $H^+$
 can still be as low as $m_{H^+} \sim 100$ GeV. The
 experimental lower bound on the mass of the lightest neutral Higgs
 boson \cite{Amsler:2008zz} implies that $\tan\beta \gtrsim 3$ in the
 MSSM.

\begin{figure}[h]
\begin{center}
\includegraphics[width=14cm]{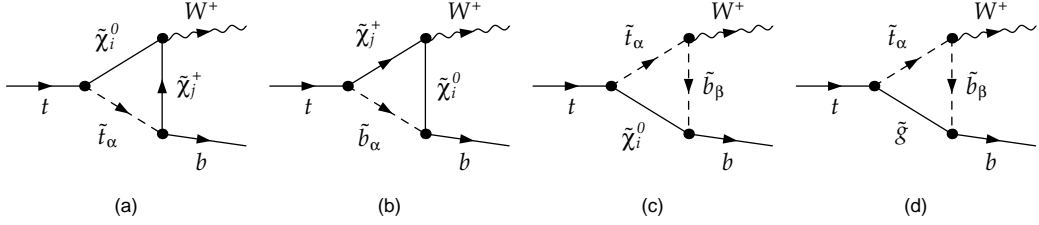}
\end{center}
\caption{Feynman diagrams for the 1-loop SUSY electroweak (a - c) and QCD
  (d) contributions
  to the $tbW$ vertex in the MSSM. The contributions to the two-point functions
  that are involved in the renormalization are not shown.}
 \label{fig:sew}
\end{figure}

The ``genuine'' one-loop MSSM corrections to the $t\to b W$ decay amplitude are
SUSY QCD corrections due to the exchange of gluinos ${\tilde g}$ and squarks
${\tilde q}_{1,2}$, and SUSY electroweak (SUSY EW) corrections which arise from
the exchange of squarks, charginos ${\tilde \chi}^\pm_i$, and neutralinos
${\tilde \chi}^0_i$.  The corresponding Feynman diagrams are depicted in
Fig.~\ref{fig:sew}. We neglect squark-mixing between different generations;
i.e., only top and bottom squarks ${\tilde t}_{1,2}$, ${\tilde b}_{1,2}$ are
taken into account. These mass eigenstates are mixtures of the respective weak
eigenstates.

%

The SUSY QCD and SUSY EW contributions to the top-width $\Gamma(t\to bW)$ were
computed in \cite{Li:1992ga,Dabelstein:1995jt,Brandenburg:2002xa,Cao:2003yk} and
\cite{Yang:1993ra,Garcia:1993rq,Cao:2003yk}, respectively.  These contributions
tend to cancel each other: while the SUSY QCD are in general negative and
typically $\lesssim 1\%$ or below in magnitude, the SUSY EW corrections are
positive in general and also $\lesssim 1\%$ for phenomenologically acceptable
SUSY particle masses.  The SUSY QCD and SUSY EW corrections to the helicity
fractions $F_{0,\mp}$ defined in Section~\ref{secaom} were calculated in
\cite{Cao:2003yk}. The corrections to these ratios are typically of the order of
$1\%$, and tend to be of opposite sign.  We have also computed these corrections
to the top-width. We compared the SUSY QCD corrections with the results of
\cite{Brandenburg:2002xa,Cao:2003yk}, with which we agree.

In the following we assume the masses of the gluinos and of the bottom squarks
${\tilde b}_{1,2}$ to be above $350$ GeV, which is suggested by the Tevatron
searches and analyses in the context of the mSUGRA scenario (see, e.g.,
\cite{Amsler:2008zz}). As to the top squarks ${\tilde t}_{1,2}$, it is not yet
excluded that one of them is lighter than the top quark. We assume that
$m_{{\tilde t}_1} \geq 100$ GeV. For the masses of the charginos and and neutralinos
we assume the lower bounds $m_{{\tilde \chi}^+} \geq 100$ GeV,
$m_{{\tilde\chi}^0} \geq 50$ GeV.

More specifically, we use the following set of SUSY parameters: 
\bea &\mu =
250\,{\rm GeV},\quad M_1 = 62\,{\rm GeV},\quad M_2 = 130\,{\rm GeV},\quad
m_{H^+} = 250\,{\rm GeV}, &\nonumber\\ 
&M_{L_i}=M_{E_i}=M_{Q_i}=M_{D_i}=M_{U_{1,2}}=400\,{\rm GeV},\quad
 M_{U_3}=250 \,{\rm GeV},&\nonumber\\ &A_{L_i}=A_{U_{1,2}}=A_{D_i}=0,\quad
 A_{U_3}=700\,{\rm GeV},&\nonumber\\ &m_{\tilde g}=350\,{\rm GeV}. &
\eea 
Here $\mu$, $M_1$, and $M_2$ are the soft SUSY breaking masses of the Higgs
potential, $M_L$ and $M_E$ are the soft masses of the left and right-handed
sleptons, $M_Q$ the soft masses of the left-handed squarks and $M_U$ and $M_D$
the soft masses of the right-handed up- and down-type squarks. We
impose the  GUT relation $M_1=5s_W^2 M_2/(3c_W^2)$. Furthermore
$m_{\tilde g}$ is the gluino mass and $i$ is a generation index. Keeping the
above parameters fixed and varying $\tan\beta$ between $5$ and $50$ causes the
physical masses and mixing angles to vary in the ranges shown in
table~\ref{tab:mssmpar}. The Higgs-boson masses and the mixing angle $\alpha$ in
table~\ref{tab:mssmpar} were calculated with {\it FeynHiggs} (version 2.6.4)  \cite{Heinemeyer:1998yj,Heinemeyer:1998np,Degrassi:2002fi,Frank:2006yh}.
\begin{table}
  \hfill
  \begin{tabular}{lrr}
    \hline\hline
                        & $\tan\beta=3$  & $\tan\beta=50$\\
    \hline
    $m_{h_0}$           & 115\,{\rm GeV} & 120\,{\rm GeV}\\
    $m_{H_0}$           & 238\,{\rm GeV} & 232\,{\rm GeV}\\
    $m_{A_0}$           & 237\,{\rm GeV} & 236\,{\rm GeV}\\
    $\alpha$            & $-0.092\pi$    & $-0.002\pi$ \\
    $m_{\chi^+_1}$      & 105\,{\rm GeV} & 115\,{\rm GeV} \\
    $m_{\chi^0_1}$      &  56\,{\rm GeV} &  60\,{\rm GeV} \\
    \hline\hline
  \end{tabular}
  \hfill
  \begin{tabular}{lrr}
    \hline\hline
                        & $\tan\beta=5$  & $\tan\beta=50$\\
    \hline
    $m_{\tilde t_1}$    & 131\,{\rm GeV} &  99\,{\rm GeV} \\
    $m_{\tilde t_2}$    & 511\,{\rm GeV} & 518\,{\rm GeV} \\
    $\theta_{\tilde t}$ & $0.19\pi$     & $0.19\pi$ \\
    $m_{\tilde b_1}$    & 395\,{\rm GeV} & 319\,{\rm GeV} \\
    $m_{\tilde b_2}$    & 409\,{\rm GeV} & 471\,{\rm GeV} \\
    $\theta_{\tilde b}$ & $-0.23\pi$     & $-0.25\pi$ \\
    \hline\hline
  \end{tabular}
  \hspace*{\fill}\par
  \caption{The values of physical  MSSM masses and mixing angles for
    $\tan\beta=5$ and $\tan\beta=50$. Mass eigenvalues are enumerated in
    ascending order. For example, $m_{\chi^0_1}$ denotes the mass of the
    lightest neutralino. All other sfermion masses lie at $400\pm5\,{\rm GeV}$
    for both values of $\tan\beta$. The angles $\theta_{\tilde t}$ and
    $\theta_{\tilde b}$ are the stop and sbottom mixing angles,
    respectively. Since we set all lepton and light (i.e. generation 1 and 2)
    quark masses to zero, all other mixing angles are zero.}
  \label{tab:mssmpar}
\end{table}

\begin{figure}[h]
\begin{center}
\includegraphics[width=12cm]{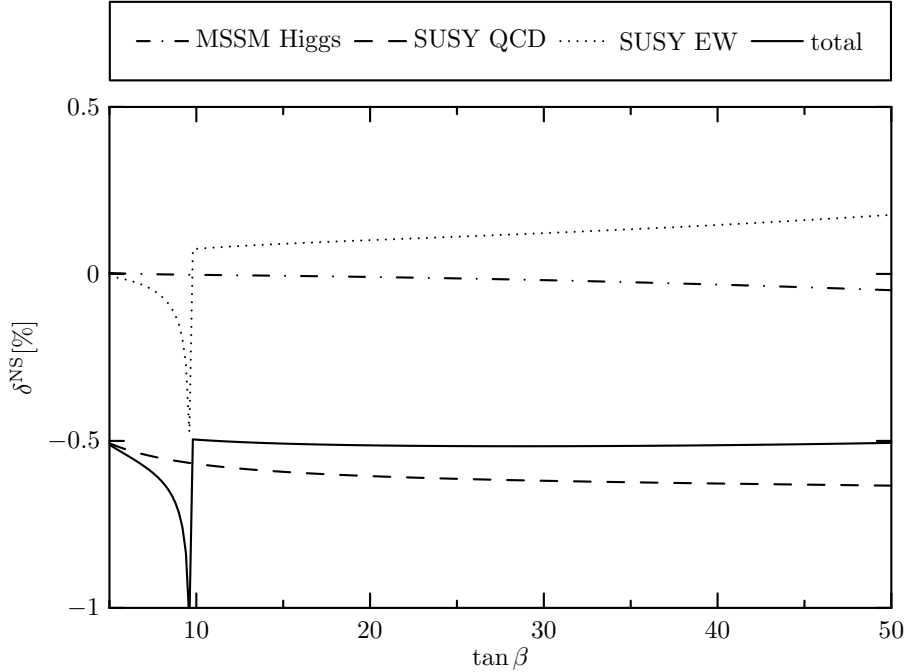}
\end{center}
\caption{The MSSM Higgs,  QCD,  and EW corrections 
$\delta_{NS}(G_F)$ to the Born width 
  as a function of $\tan\beta$. The other parameters are as given in
  table~\ref{tab:mssmpar}.}
 \label{fig:delsusy}
 \end{figure}

In Fig.~\ref{fig:delsusy} the MSSM Higgs, QCD, and EW corrections
$\delta_{NS}(G_F) =(\Gamma_{NS} - \Gamma_B(G_F))/\Gamma_B(G_F)$ and their sum
are shown as a function of $\tan\beta$. For the parameters specified above, the
MSSM Higgs contributions are very small. The SUSY EW contributions
have a sharp
peak at $\tan\beta\approx 10$, which corresponds to a threshold effect. Here
the masses are such that a top quark can decay into an on-shell neutralino and an
on-shell stop. Obviously, our results are unreliable in the vicinity of that
peak. Otherwise, the MSSM corrections are dominated by the SUSY QCD
contributions, which remain almost constant at about $-0.6\%$ for
$\tan\beta>10$. In this range, the SUSY EW contributions yield a constant
$+0.1\%$, leading to an overall correction of $-0.5\%$.

The bulk of the corrections $\delta_{NS}(G_F)$ is again due to the renormalized
form factor $f_L$.  For the parameters given above we found the following
features: $|f_L|> |g_R| \gg |f_R|, |g_L|$ and $|{\rm Re} \, f_L| \,
, |{\rm Re}
\,g_R| \gg |{\rm Im} \, f_L|, |{\rm Im} \,g_R|$. Therefore we display
  only ${\rm Re} \, f_L$ and ${\rm Re} \,g_R$.

\begin{figure}
\begin{center}
\includegraphics[width=12cm]{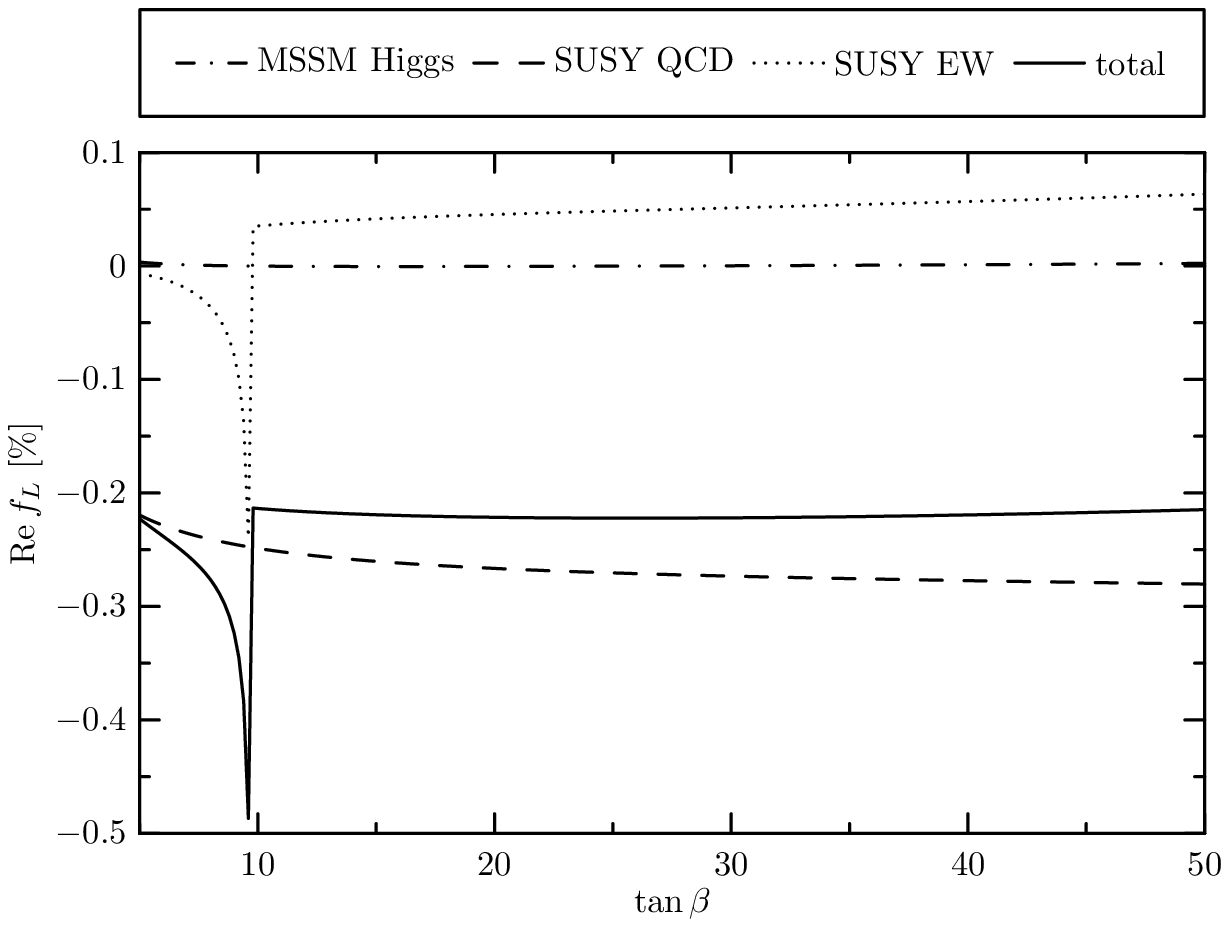}
\end{center}
\caption{The anomalous form factor ${\rm Re} \, f_L$ induced by the 
 MSSM  Higgs, QCD, and EW corrections  as a function of $\tan\beta$.
 The other parameters are as given in table~\ref{tab:mssmpar}.}
 \label{fig:reflsusy}
 \end{figure}

\begin{figure}
\begin{center}
\includegraphics[width=12cm]{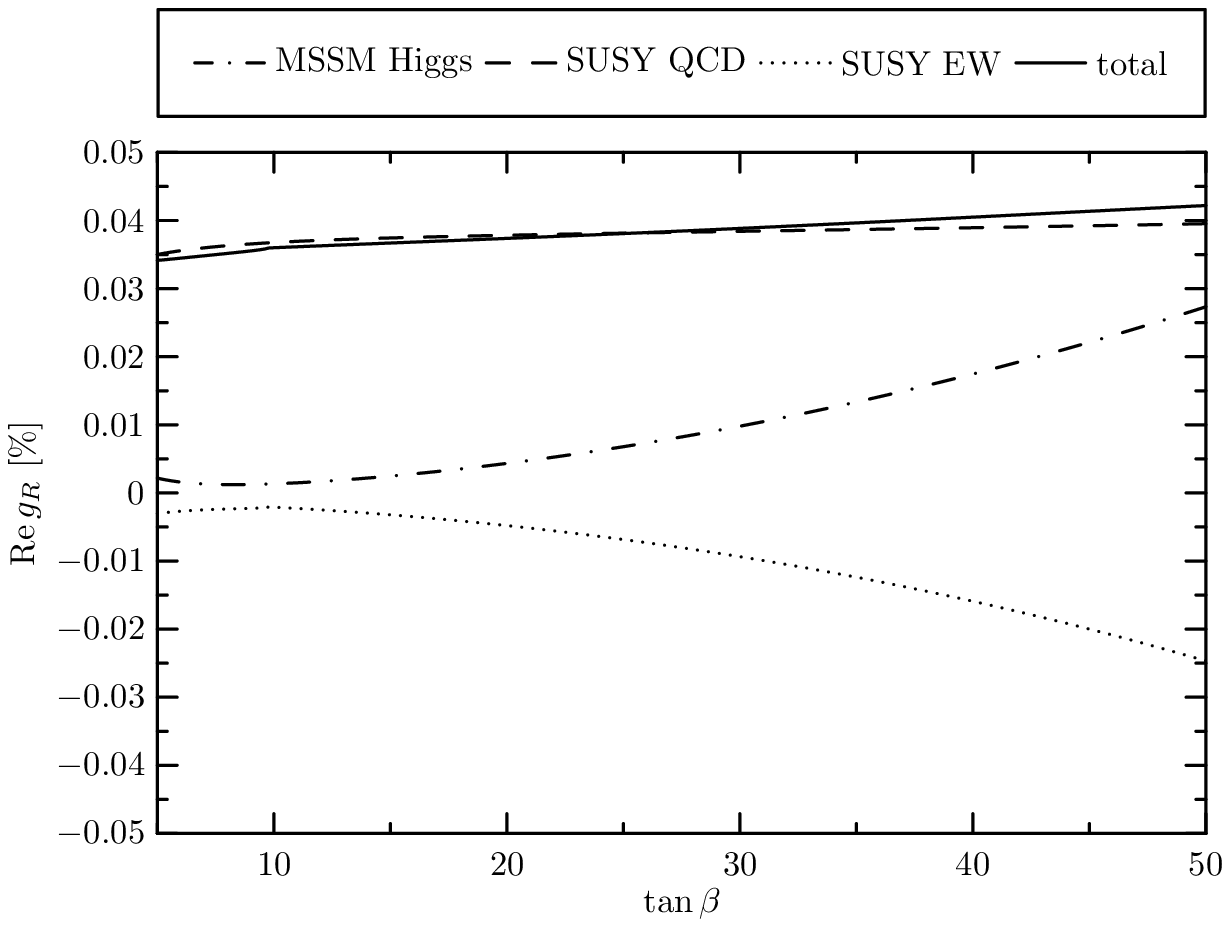}
\end{center}
\caption{The anomalous form factor ${\rm Re} \, g_R$ induced by the MSSM Higgs,
  QCD, and EW corrections as a function of $\tan\beta$.  The other parameters
  are as given in table~\ref{tab:mssmpar}.}
 \label{fig:regrsusy}
 \end{figure}

In Figs.~\ref{fig:reflsusy} and~\ref{fig:regrsusy} the real parts of
$f_L$ and $g_R$ induced by the various MSSM
corrections are shown as a function of $\tan\beta$ for the above parameter set.
In the latter case, the Higgs and SUSY EW corrections cancel almost
exactly. Thus  ${\rm Re} g_R$ and ${\rm Re} f_L$ are essentially due to the SUSY QCD
contribution.

For completeness, we briefly address also the effect of supersymmetric
$CP$-violating phases. As is well known, many new $CP$ phases can be
present in the MSSM in general. In order to assess the  size of SUSY
 $CP$ violation it is useful to consider a simplified scenario where,
 in a certain phase convention, observable $CP$ phases reside only in 
 the $\mu$
 parameter of the bilinear term in the Higgs superfields and in the
 trilinear couplings $A_f$. The experimental upper bounds on the
 electric dipole moments of the electron, the neutron, and certain
 atoms constrain the phases of  $\mu$ and of  $A_f$ for the first two
  generations to very 
small values (see, for instance, \cite{Pospelov:2005pr}). However, 
 the phases of $A_{t,b}$ may be of order one. 
 This generates $CP$-violating interactions
 between gauginos and quarks and
 squarks of the third generation\footnote{In addition, neutral 
  Higgs sector $CP$ violation is induced
  at the 1-loop level, which may be sizeable
  \cite{Pilaftsis:1998dd}.}.
Then the  imaginary parts of the anomalous form factors receive
 also contributions induced by the $CP$ phases ${\rm
   arg}(A_{t,b})$,  which in turn generate a non-zero
 triple-correlation
 asymmetry $\langle{\cal O}\rangle -  \langle{\bar{\cal O}}\rangle$, 
  discussed in  Section~\ref{secaom},
or a difference  $A_{CP}$ in the partial rates  
 (cf. eq. (\ref{eq-aspardr})).
Such effects were discussed before in 
 \cite{Bernreuther:1993xp,
   Grzadkowski:1993gh,Christova:1993jx,Aoki:1998gsa}.
 For SUSY particle masses used above, these effects are below $0.1\%$.
\clearpage

\subsubsection{Top-color assisted technicolor and Little Higgs models}
\label{substc2}
The concept that electroweak symmetry breaking (EWSB) and the generation of
quark and lepton masses occur ``dynamically'' by the condensation of (new)
fermion-antifermion pairs is still an alternative to the Higgs mechanism not
ruled out by experiment. Among the phenomenologically acceptable models that use
this concept is top-color assisted technicolor (TC2)
\cite{Hill:1994hp,Buchalla:1995dp}. TC2 has two strongly interacting sectors in
order to explain EWSB and the large top-quark mass. Technicolor interactions
(TC) are responsible, via the condensation of techni-fermions, $\langle {\bar T}
T\rangle$ ($T = U, D)$, for most of EWSB, but they contribute very little to the
top-quark mass $m_t$, while top-color interactions (TopC) generate through
condensation of top quarks, $\langle {\bar t} t \rangle$, the bulk of $m_t$ but
make only a small contribution to EWSB.  The spin-zero states of the model are
bound-states of the techni-fermions and of $t, b$. These two sets of
bound-states form two $SU(2)_L$ doublets $\Phi_{TC}, \Phi _t$, whose couplings
to the weak gauge bosons and to $t$ and $b$ are formally equivalent to a
two-Higgs doublet model. The physical spin-zero states are i) a heavy neutral
scalar $H_{TC}$ with a mass of order 1 TeV, ii) a neutral scalar $H_t$ which is
a ${\bar t} t$ bound state whose mass is expected to be of the order $m_{H_t}
\sim 2 m_t$ when estimated \`a la Nambu-Jona-Lasinio, but could also be
considerably lighter \cite{Chivukula:1998wd}, and iii) a neutral
``top-pion'' $\Pi^0$ and a pair of
charged ones, $\Pi^\pm$, whose masses are predicted to lie in the range
of 180 - 250 GeV \cite{Hill:1994hp,Buchalla:1995dp}. Below we shall use $m_{H_t}
\geq 120$ GeV, $m_{\Pi^0}= m_{\Pi^+} \geq 180$ GeV, and $m_{H_{TC}} = 1$ TeV.

The Yukawa couplings of the top quark to the physical spin-zero states read
after EWSB \cite{Leibovich:2001ev}:
\begin{eqnarray} \label{yuk-tc2}
{\cal L}_{Y} = 
 - \frac{1}{\sqrt 2} (Y_t f_\pi + \epsilon_t v_T) {\bar
  t} t
- \frac{1}{\sqrt 2} (Y_t H_t + \epsilon_t H_{TC}){\bar
  t} t \nonumber \\
 - (i \frac{Y_\pi}{\sqrt 2} \Pi^0 {\bar t}_L t_R
  + i  Y_\pi  \Pi^- {\bar b}_L t_R + {\rm h.c.} ) \, ,
\end{eqnarray}
where $Y_\pi =(Y_t v_T -\epsilon_t f_\pi)/v$.  Here $f_\pi$ denotes the value of
the top-quark condensate which is estimated in the TC2 model to be $f_\pi \sim$
60 GeV \cite{Hill:1994hp,Leibovich:2001ev}. Once $f_\pi$ is fixed, $v_T$ is
determined by the EWSB requirement that $f_\pi^2 + v_T^2 = v^2 = (246 \, {\rm
  GeV})^2$. From (\ref{yuk-tc2}) one sees that $(Y_t f_\pi + \epsilon_t
v_T)/\sqrt{2} = m_t.$ The technicolor contribution $\epsilon_t$ to the top mass
is small, by construction of the TC2 model. We have therefore set $\epsilon_t=0$
in our calculation. As a consequence the top Yukawa coupling $Y_t$ becomes
large, i.e.\ $Y_t\simeq 4$. Therefore, the top quark is expected to couple
strongly to $H_t$ and to the top-pions. The coupling of the charged top-pion to
$b_R$ is very small, and likewise the couplings of $\Pi^0$ and $H_t$ to $b$
quarks. We shall therefore neglect them below.  The remaining interactions of the two
doublets $\Phi_{TC}$, $\Phi _t$ with $t, b$ and with the weak gauge bosons can
be found in \cite{Leibovich:2001ev} whose conventions we use here.

The 1-loop new physics contributions in the TC2 model to the $t \to b W$ decay
amplitude correspond to the diagrams of  Fig.~\ref{fig:higgs} with the replacements
$h^0 \to H_{TC}$, $H^0 \to H_t$, $A^0 \to \Pi^0$, and $H^\pm \to \Pi^\pm.$ In
addition there are the 1-loop contributions of these spin-zero states to the
wave-function renormalization constants and to $\Delta r$. As $H_{TC}$ is very
heavy and its Yukawa couplings to $t$ and $b$ are small, the contributions of
the corresponding diagrams are suppressed with respect to the remaining
corrections. The same remark applies to the diagrams where $\Pi^0$ and $H_t$
couple to $b$ quarks. A closer inspection shows that, for relatively light
$\Pi^{0,\pm}$ and $H_t$ the two dominant contributions are Fig.~\ref{fig:higgs}
(b) with $\phi_i = H_t$ and $\phi_i = \Pi^0$ while Figs.~\ref{fig:higgs}(d) and
(f) are subdominant.

The corrections to the top-width and to the helicity fractions 
 were computed in \cite{Wang:2005ra}, in the conventional on-shell
renormalization scheme ($\alpha_{em}$ scheme, for a TC2 model
slightly different from that outlined in \cite{Leibovich:2001ev}. 
 Using the couplings and masses of \cite{Wang:2005ra} 
 we find agreement with this this paper.

\begin{figure}
\begin{center}
\includegraphics[width=12cm]{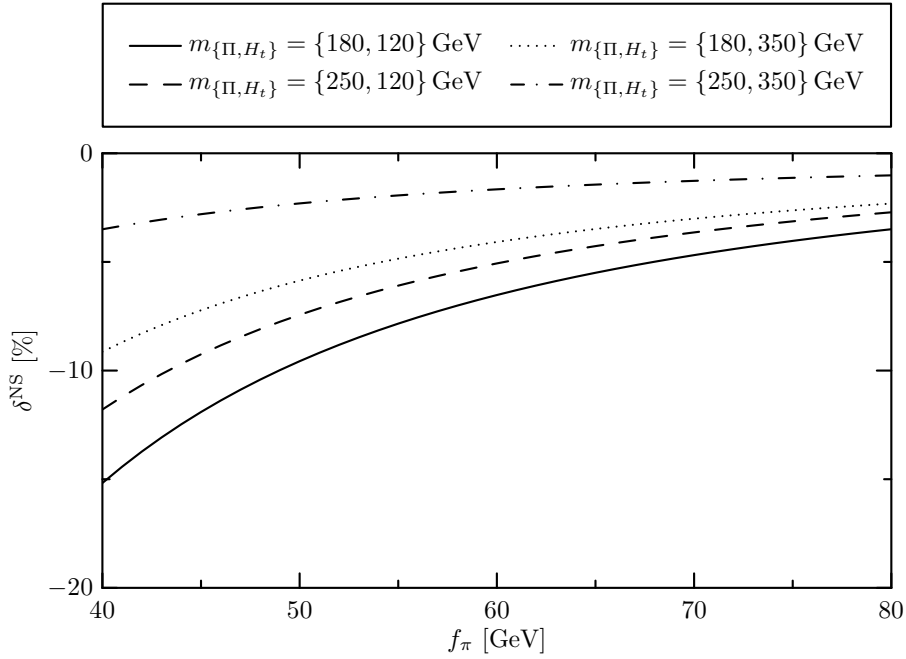}
\end{center}
\caption{The correction $\delta_{NS}(G_F)$ to the Born width in the TC2 model as
  a function of $f_\pi$.}
 \label{fig:deltc2}
 \end{figure}

Fig.~\ref{fig:deltc2} shows the correction $\delta_{NS}(G_F) =(\Gamma_{NS} -
\Gamma_B(G_F))/\Gamma_B(G_F)$ as a function of the value of the top-quark
condensate $f_\pi$, for various sets of masses $m_{H_t},$ $m_\Pi$ of the
top-Higgs boson and top-pions.  For fixed $m_{H_t},$ $m_\Pi$ the corrections
increase in magnitude with increasing top-Yukawa coupling, i.e., decreasing $f_\pi$. As
mentioned above, the dominant contributions are the ones corresponding to
Fig.~\ref{fig:higgs} (b) with $\phi_i = H_t$ and $\phi_i = \Pi^0$. The
correction  $\delta_{NS}(G_F)$ is negative in the above parameter
range. It can become as large as  $\sim - 15\%$. 

\begin{figure}
\begin{center}
\includegraphics[width=12cm]{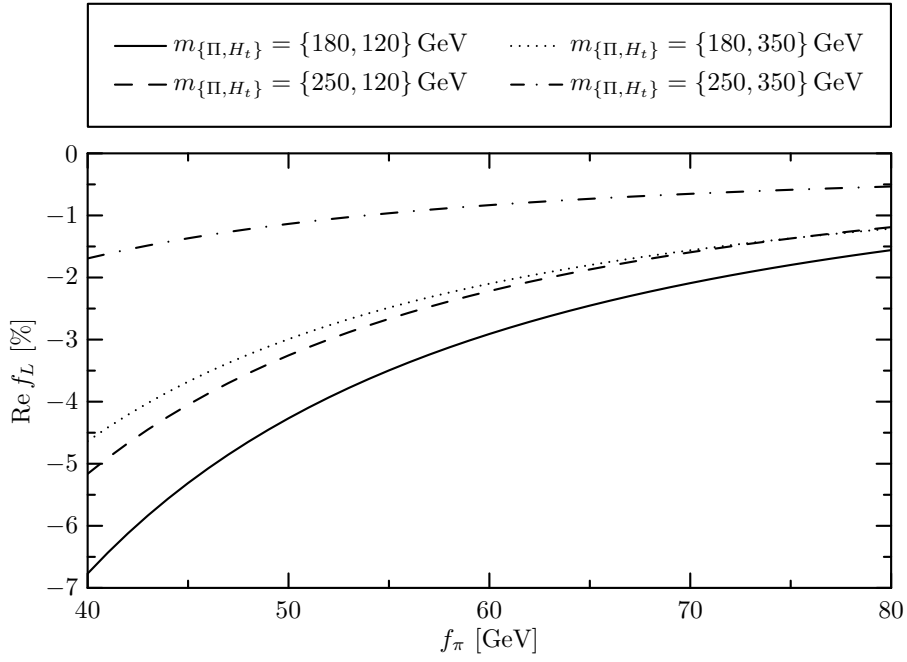}
\end{center}
\caption{The anomalous form factor ${\rm Re} \, f_L$ in the TC2 model
  as a function of $f_\pi$.}
 \label{fig:refltc2}
 \end{figure}

\begin{figure}
\begin{center}
\includegraphics[width=12cm]{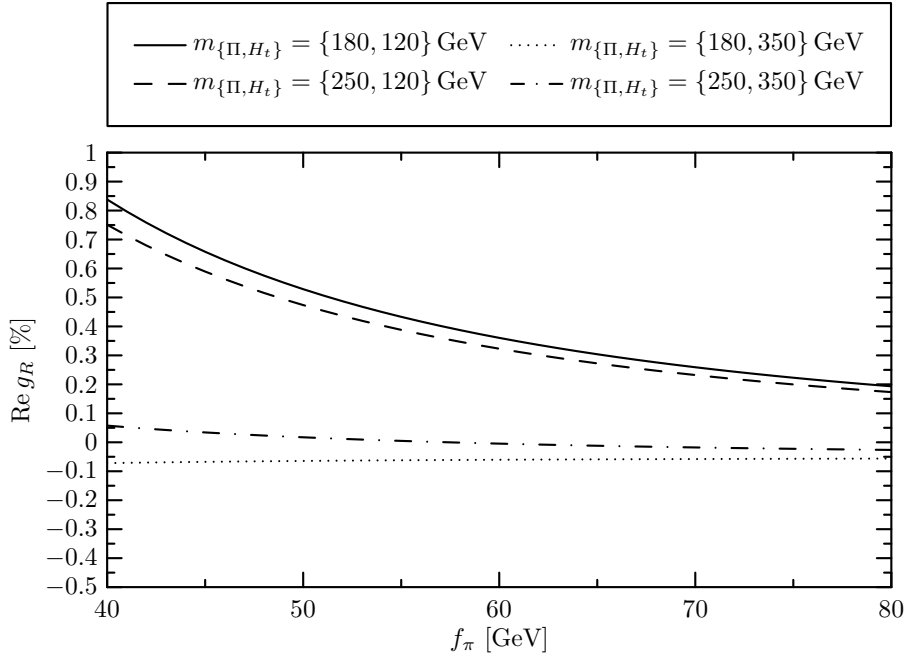}
\end{center}
\caption{The anomalous form factor ${\rm Re} \, g_R$  in the TC2 model
  as a function of $f_\pi$.}
 \label{fig:regrtc2}
 \end{figure}

The renormalized form factor ${\rm Re} \, f_L$ is plotted in
Fig.~\ref{fig:refltc2} as a function of $f_\pi$. As expected, ${\rm Re} \, f_L
\approx \delta_{NS}/2$.  The chirality-flipping form factor ${\rm Re} \, g_R$ is
shown in Fig.~\ref{fig:regrtc2}. The magnitude of this form factor remains
below the percent level also in this model.

For completeness, we mention  some results from Little Higgs models. These
models incorporate heavy partners (with TeV scale masses) of the weak gauge
bosons and of the top quark. An often studied version of these models is the
``Littlest Higgs'' model (LH) \cite{ArkaniHamed:2002qy} with $T$-parity symmetry
\cite{Cheng:2003ju}. In this model, the SM $tbW$ vertex is modified already at
tree-level due to the mixing of $t_L$ with the heavy top-quark partner $T$ and the
mixing of the weak gauge bosons $W^\pm$ with their heavy partners $W_H^\pm$. The
resulting anomalous coupling $f_L$ and  the correction to the top-quark
decay width $\delta \Gamma_t/\Gamma_t$ are negative; the latter can become larger
than $10\%$ in magnitude \cite{Cao:2006wk,Berger:2005ht}. One-loop LH radiative
corrections induce also a coupling ${\rm Re} \, g_R$ which is, however, too
small to be observable at the LHC \cite{Penunuri:2008pb}.

\section{Conclusions}
\label{sec-conc}

In this paper we have analyzed the magnitudes and phases of the 
anomalous form factors $f_{L,R}$ and $g_{L,R}$ in the  $tWb$
vertex within several SM extensions,
 to wit, a 2HDM, the MSSM, and a TC2 model.  We found that the 
imaginary parts of the form factors, which can be induced either by 
$CP$-invariant final-state rescattering or by $CP$-violating
interactions, are very small compared to the real parts. Moreover,
within the above models,  $|{\rm Re} f_R|, |{\rm Re} g_L| \ll|{\rm Re}
g_R| <|{\rm Re} f_L|$. In the 2HDM and the MSSM, where electroweak
symmetry breaking is triggered by elementary Higgs fields, the magnitudes
of the anomalous couplings $f_L$, $g_R$ are smaller than $1\%$.
TC2 and Little Higgs models are viable paradigms for the special role the top
can play in the mechanism of electroweak symmetry breaking. TC2
interactions may reduce $f_L$
significantly, which would reduce the top width by $10\%$ or more. A
 reduction of similar size can happen in Little Higgs models. In the
 long run, this can be tested in single-top-quark production at the
 LHC, where one may eventually measure  $f_L$ with a
 precision of about $5\%$. The determination  of the  top width
 $\Gamma_t$ with an
 accuracy of about $10\%$ would require a high-energy $e^+e^-$ linear
 collider \cite{AguilarSaavedra:2001rg}, where 
  $\Gamma_t$ could be obtained from a precision measurement 
 of  $t \bar t$ production at threshold.

\subsubsection*{Acknowledgments}
We wish to thank Peter Zerwas for useful discussions 
and J.~j.~Cao for a correspondence concerning Ref. \cite{Cao:2003yk}.
This work was supported by Deutsche Forschungsgemeinschaft 
 SFB/TR9.


\end{document}